\documentclass{article}       
\usepackage{times}
\usepackage{amsmath,amsfonts,amssymb}
\usepackage{graphicx}

\begin{document}              
\pagenumbering{arabic} \baselineskip 15pt \date{}
\pagestyle{myheadings} \markboth{\footnotesize}{\footnotesize Risk-Limiting Bayesian Polling Audits for Two Candidate Elections}
 \title{Risk-Limiting Bayesian Polling Audits for Two Candidate Elections}
 \author{Poorvi  L. Vora\\Department of Computer Science, The George Washington University}
\date{\today}       
\maketitle                    

\begin{abstract}
We propose a simple common framework for Risk-Limiting and Bayesian (polling) audits for two-candidate plurality elections. Using it, we derive an expression for the general Bayesian audit; in particular, we do not restrict the prior to a beta distribution. We observe that the decision rule for the Bayesian audit is a simple comparison test, which enables the use of pre-computation---without simulations---and greatly increases the computational efficiency of the audit. Our main contribution is a general form for an audit that is both Bayesian and risk-limiting: the {\em Bayesian Risk-Limiting Audit}, which enables the use of a Bayesian approach to explore more efficient Risk-Limiting Audits.
\end{abstract}
\section{Introduction}
\label{sec:intro}
The framework of {\em risk-limiting audits (RLAs)}, as described by Lindeman and Stark \cite{RLA}, formalizes a rigorous approach to election verification. The purpose of an audit is to require a full hand count if the outcome is wrong; the {\em risk} is the rate at which it fails to do so. A {\em RLA} is one that guarantees that the risk will be smaller than a pre-specified bound.

Computing the risk is tricky, however. If the audit rule and parameters are fixed, the risk depends on the (unknown) true tally of the election. Smaller margins in favor of the declared loser make it harder to detect that the outcome is wrong, and correspond to larger values of the risk. An audit typically uses a single measure to represent the varied values the risk can take as the underlying (unknown) tally varies.
\begin{itemize}
\item A {\em RLA} uses as its measure the largest possible risk, corresponding to the wrong outcome which is hardest to detect: either a tie or a margin of one vote in favor of the(a) declared loser.
\item The Bayesian audit, as described by Rivest and Shen \cite{bayesian-audits}, is a newer approach to election audits. It considers each possible margin in favor of the declared loser(s). The Bayesian measure, termed the {\em upset probability}, is a weighted average of the corresponding risks.
\end{itemize}
Each type of audit samples ballots, stopping when the risk measure is smaller than a guaranteed upper bound.

Depending on what the true tally is, there could be considerable difference between the measure, which is computed from the sampled ballots, and the true risk, which is unknown. While all we can compute is the measure, it is the true risk we care about.
\begin{itemize}
\item The true risk is never larger than the {\em RLA} measure (maximum risk). The {\em RLA} stops when the maximum risk is smaller than the bound; the true risk will also be smaller than the bound.
\item The true risk of a close election could be larger than the Bayesian measure, the upset probability, which is an average risk. The Bayesian audit stops when the upset probability is smaller than the bound, but the true risk might still be too large. {\em There is no way for the observer to know whether this is the case, because the true tally is unknown}.
\end{itemize}

In addition to proposing the Bayesian approach, Rivest and Shen also propose the use of P\'{o}lya urn simulations to compute whether an audit should end or not. While the simulations are not efficient enough for real time use, they provide the only way we know to carry out the audit of a complex election. Traditional {\em RLA}s are more efficient for plurality elections and can be reduced to a comparison test as described in the classical work of Wald \cite{wald}. The CLIP audit of Rivest \cite{clip} is another {\em RLA} which may also be reduced to such a comparison, though the values are computed using simulations\footnote{Philip Stark has mentioned work in progress: a CLIP-like audit which does not use simulations.}.

The Bayesian framework is exceptionally promising as a means of designing efficient audits (requiring a small sample size). As we have seen above, important open research areas include (a) the characterization of the risk limit (or maximum risk) of a Bayesian audit and (b) improving its computational efficiency.

The following are important open questions regarding characterization of Bayesian audits: Is there a well-defined relationship between {\em RLA}s and Bayesian audits? What is the relationship between the largest risk ({\em RLA} risk measure) and the upset probability (Bayesian audit risk measure)? What is the closest election for which the upset probability is not smaller than the risk? That is, what is the closest election for which the upset probability is a reasonable stand-in for the true risk? Can Bayesian audits be designed to be comparable to traditional {\em RLA}s in computational efficiency? Also of interest are questions regarding audits that are both Bayesian and risk-limiting: What form might Bayesian risk-limiting audits ({\em RLA}s)---where the upset probability is identical to the largest risk---take? Could Bayesian {\em RLA}s be designed to use a smaller sample than traditional {\em RLA}s?

While we do not attempt to answer all the questions posed above, we present early results that should provide a basis for exploring the answers. In this paper, while restricting ourselves to a two-candidate election and polling audits, we view both audits in a single framework. Among Bayesian audits, we study only those with equal prior winning probabilities for each candidate. We expect that our results will be applicable in a straightforward fashion to comparison audits for two-candidate plurality elections as well, though that is work in progress.
\subsection{Our contributions}
Our contributions are as follow:
\begin{enumerate}
\item
We show that the Bayesian audit can be reduced to a simple comparison test between the number of votes for the winner in the audit sample and two pre-computed values for this sample size:
\begin{itemize}
\item a minimum number of votes for the winner, $k^+$, above which the election outcome is declared correct, and
\item a maximum number of votes for the winner, $k^-$, below which the audit proceeds to a hand count.
\end{itemize}

The values of $k^+$ and $k^-$ can be pre-computed, making the Bayesian audit a feasible real-time audit, as it is a simple look-up process at the time of drawing samples. If Bayesian audits for more complex elections---such as those with multiple candidates (we are still working on this)---could also be reduced to comparison tests, computational efficiency could be greatly improved.

\item We show that the traditional {\em RLA} as described in \cite{RLA} is a Bayesian audit with a prior that assumes that, if the election were correct, the announced winner would earn a fraction $p$ of the votes; if not, she would win half the votes. Such an incorrect outcome is the hardest possible to distinguish from a correct one, which is what makes the audit a {\em RLA}.

\item Motivated by our understanding of the traditional {\em RLA}, and restricting ourselves to odd $N$, we define a class of Bayesian audits that are also {\em RLA}s. An audit in this class may have any prior on tally values for a correct outcome, but the only allowed tally for an incorrect outcome is that corresponding to the closest possible election lost by the winner (in this case, a margin of one). We prove that this is a {\em RLA}. For such a prior, we show that the upset probability of the Bayesian audit is also the largest risk.

\item We report on verification of:
\begin{enumerate}
\item our comparison test for the Bayesian audit (1)
\item the risk limit of the audit in (3).
\end{enumerate}
\end{enumerate}

We draw considerably from the proof approaches in \cite{wald}.

\subsection{Organization}
This paper is organized as follows. Section \ref{sec:model} describes the model and establishes most of the notation. Section \ref{sec:review} describes the Wald Sequential Test \cite{wald}, {\em RLA}s \cite{RLA} and Bayesian audits \cite{bayesian-audits}. Our contributions are to be found in sections \ref{sec:theorems} and \ref{sec:expts}. Section \ref{sec:theorems} presents our results on a simple form of the Bayesian audit and a new audit that is both Bayesian and risk-limiting, the Bayesian {\em RLA}. Section \ref{sec:expts} presents look-up table forms of the audits and verification of the simple form of the Bayesian audit and the risk of a Bayesian {\em RLA}. Section \ref{sec:conc} concludes. Acknowledgements are in section \ref{sec:ack} and proofs in the Appendix.
\section{The Model}
\label{sec:model}
We consider a plurality election with two candidates, $N$ voters and no invalid votes. We assume that $N$ is odd so the winner is well-defined. It is not hard to extend the results to even $N$, we use odd $N$ for ease of exposition. We denote by $W$ the random variable representing the true winner, by $w$ an instance of $W$, by $w_a$ and $\ell _a$ the announced winner and loser respectively and by $x$ the (true, unknown) number of votes obtained by $w_a$. Thus $w_a=w$ if and only if $x > \frac{N}{2}$.

A polling audit will estimate whether $w_a$ is the true winner. Consider a sample of $n$ votes drawn uniformly at random: $v_1$, $v_2$, ..., $v_n$, $n < N$, $v_i \in \{w_a, \ell_a\}$. The sample forms the {\em signal} or the {\em observation}; the corresponding random variable is denoted ${\bf S}_n \in \{w_a, \ell_a\}^n$, the specific value ${\bf s}_n = [v_1, v_2, ..., v_n] $. Let $k_{{\bf s}_n}$ denote the number of votes for $w_a$ in the sample; then $n-k_{{\bf s}_n}$ votes are for $\ell _a$.

The audit computes a binary-valued estimate of the true winner from ${\bf s}_n$: \[\hat{w}_n: \{w_a, \ell_a\}^n \rightarrow \{w_a, \ell_a\} \] We will refer to the function $\hat{w}_n$ as the {\em estimator} and $\hat{w}_n({\bf s}_n)$ as the {\em estimate}. The audit uses an error measure to compute the quality of the estimate.
\begin{itemize}
\item If $\hat{w}_n({\bf s}_n)=w_a$ and the error measure is acceptable we are done (stop) and declare that the election outcome was correctly announced.
\item If $\hat{w}_n({\bf s}_n) = \ell_a$ and the error measure is acceptable we stop drawing votes and proceed to perform a complete hand count.
\item If the error measure is not acceptable we draw more votes to improve the estimate.
\end{itemize}

In computing $\hat{w}_n({\bf s}_n)$, we can make two types of errors:
\begin{enumerate}
\item {\em Miss:} A {\em miss} occurs when the announced outcome is incorrect, $w \neq w_a$, but the audit misses this, and $\hat{w}_n({\bf s}_n)=w_a$. We denote by $P_M$ the probability of a miss---given that the announced outcome is incorrect, the probability that the audit will miss this: \[ P_M = Pr[\hat{w}_n({\bf s}_n)=w_a \mid w \neq w_a] \]
$P_M$ is the {\em risk} in risk limiting audits. If the audit is viewed as a statistical test, with the null hypothesis being $w=\ell_a$, $P_M$ is the Type I error.
\item {\em Unnecessary Hand Count:} Similarly, if $w=w_a$, but $\hat{w}_n({\bf s}_n) = \ell_a$, acceptance of the estimate would lead to an unnecessary hand count. We denote the probability of an {\em unnecessary hand count} by $P_U$: \[P_U = Pr[\hat{w}_n({\bf s}_n) = \ell_a \mid w=w_a] \] If the audit is viewed as a statistical test, with the null hypothesis being $w=\ell_a$, $P_U$ is the Type II error.
\end{enumerate}

\section{Defining the audit}
\label{sec:review}
In this section, we describe three types of audits. We do not attempt to introduce any new ideas, but try to faithfully represent the existing literature.

\begin{enumerate}
\item {\em Wald Sequential Tests}

A classical approach is the Wald Sequential Test which limits both $P_M$ and $P_U$. \\

{\em Definition 1:} The {\em Wald Sequential Test} is the {\em likelihood ratio test} \cite{wald}:
\begin{equation}
\hat{w}_n =  \left\{ \begin{array}{ll} w_a & ~~~\sigma_n > \frac{1-\beta}{\alpha}\\
& \\
\ell _a& ~~~ \sigma_n < \frac{\beta}{1-\alpha} \\
& \\
undetermined ~(draw~more~samples) & ~~~else \\
\end{array}
\right .
\label{eqn:wald-test}
\end{equation}
where $\sigma _n$ is the {\em likelihood ratio}:

\[ \sigma _n =  \frac{Pr[{\bf S}_n = {\bf s}_n \mid w=w_a]}{Pr[{\bf S}_n = {\bf s}_n \mid w = \ell_a] }\]

and $0 \leq \alpha, \beta < \frac{1}{2}$. \\

{Proposition 1 \cite{wald}:} The Wald Sequential Test has $P_M < \alpha$ and $P_U < \beta$, and is a most efficient test achieving these bounds.\\

(A {\em most efficient} test is one requiring the smallest sample size). An argument supporting Proposition 1 may be found in \cite{wald}. \\

Suppose the draws are independent (with replacement) and $k_{{\bf s}_n}$ of the $n$ votes in the sample are for $w_a$. To compute the expressions $Pr[{\bf S}_n = {\bf s}_n \mid w=w_a]$ and $Pr[{\bf S}_n = {\bf s}_n \mid w = \ell_a]$ required by the test, we need $x$, the election's true vote counts for $w_a$, when $w=w_a$ and $w = \ell _a$ respectively. Assume that if $w_a$ wins, she wins with $p_1N$ (greater than half) votes, and if she does not ($\ell _a$ wins or the election is a tie), $w_a$ obtains $p_0N$ (no more than half) votes, where $p_1N$ and $p_0N$ are integers, and $\frac{1}{2} < p_1 \leq 1$ and $0 \leq p_0 \leq \frac{1}{2}$. With these assumptions, the Wald Sequential Test is easily seen to be \cite{wald}:
\begin{equation}
\hat{w}_n =  \left\{ \begin{array}{ll} w_a & ~~~\frac{p_1^{k_{{\bf s}_n}} (1-p_1)^{n-k_{{\bf s}_n}}}{p_0^{k_{{\bf s}_n}} (1-p_0)^{n-k_{{\bf s}_n}}} > \frac{1-\beta}{\alpha}\\
& \\
\ell _a& ~~~ \frac{p_1^k (1-p_1)^{n-k_{{\bf s}_n}}}{p_0^{k_{{\bf s}_n}} (1-p_0)^{n-k_{{\bf s}_n}}} < \frac{\beta}{1-\alpha} \\
& \\
undetermined ~(draw~more~samples) & ~~~else \\ \\
\end{array}
\right .
\label{eqn:wald-p0-p1}
\end{equation}

We will refer to the test defined by (\ref{eqn:wald-p0-p1}) as the $(\alpha, \beta, p_0, p_1)$ Wald Sequential Test.

\noindent {\em Corollary 1:} When the only possible values of the true vote count, $x$, are $p_1N$ ($w_a$ wins) or $p_0N$ ($w_a$ loses), the $(\alpha, \beta, p_0, p_1)$ Wald Sequential Test has $P_M < \alpha$ and $P_U < \beta$, and is a most efficient test achieving these bounds.

\noindent {\em Proof:} This follows from Proposition 1. \\

\item {\em Risk-Limiting Audits (RLAs)} \cite{RLA}

The Wald Sequential Test requires prior knowledge of $p_1N$ and $p_0N$, the values of $x$ when $w_a$ is the winner and loser respectively. Unless one performs a full hand count, however, one does not know the true value of $x$. If $p_0$ and $p_1$ are misestimated, the true upper bounds on the risk and the probability of an unnecessary recount might not be $\alpha$ and $\beta$ respectively.

This is particularly important because the audit outcome $\hat{w}_n({\bf s}_n) = w_a$ is final (while the outcome $\hat{w}_n({\bf s}_n) = \ell _a$ is followed by a confirmatory full hand count). At the very least then, we should guarantee an upper bound on worst case errors when the audit outcome is $\hat{w}_n({\bf s}_n) = w_a$. That is, we would like to bound the risk, independent of the true value of $x$.

A {\em risk-limiting audit (RLA)} with {\em risk limit} $\alpha$---as described by, for example, Lindeman and Stark \cite{RLA}---is one for which the risk is smaller than $\alpha$ for all possible (unknown) true tallies in the election (or---equivalently for the two-candidate election---all possible values of $x$). For convenience when we compare audits, we refer to this audit as an $\alpha$-{\em RLA}.

There are many functions $\hat{w}_n$ that would satisfy the $\alpha$-{\em RLA} criterion, and not all would be desirable. For example, the constant estimate $\hat{w}_n({\bf s}_n) = \ell_a$ always requires a hand count and is risk-limiting with $P_M=0 < \alpha$, $\forall \alpha$, $\forall x$. However, $P_U=1$, and the average number of votes examined by the audit is $N$; this is undesirable.

A more efficient example of an $\alpha$-{\em RLA} is the traditional {\em RLA} \cite{RLA} based on Wald sequential tests:
\begin{equation}
\hat{w}_n =  \left\{ \begin{array}{ll} w_a & ~~~\frac{p^{k_{{\bf s}_n}} (1-p)^{n-k_{{\bf s}_n}}}{(\frac{1}{2})^n} > \frac{1-\beta}{\alpha}\\
& \\
\ell _a& ~~~ \frac{p^{k_{{\bf s}_n}} (1-p)^{n-k_{{\bf s}_n}}}{(\frac{1}{2})^n} < \frac{\beta}{1-\alpha} \\
& \\
undetermined ~(draw~more~samples) & ~~~else \\ \\
\end{array}
\right .
\label{eqn:bravo}
\end{equation}
where $p$ depends on the fraction of votes declared for $w_a$ and $\beta$ is the desired upper bound on $P_U$. We denote this the $(\alpha, \beta, p)$-{\em traditional RLA} and note that it is identical to the $(\alpha, \beta, \frac{1}{2}, p)$ Wald Sequential Test. When $\beta=0$, this is the {\em BRAVO} audit \cite{bravo}, which may be denoted the $(\alpha, p)$-{\em BRAVO} audit.

Other {\em RLA}s include the CLIP audit \cite{clip} which may be expressed as a simple comparison test between the number of votes for the winner and a pre-computed value that depends on sample size. \\

\item {\em Bayesian Audits} \cite{bayesian-audits}

Bayesian audits, defined by Rivest and Shen \cite{bayesian-audits}, assume knowledge of a {\em prior} probability distribution on $x$; we denote this distribution by $f_X$. In this model, the variable $W$ inherits a prior distribution from $f_X$, because \[ Pr[w=w_a] = \sum_{x=\frac{N+1}{2}}^N f_X(x) \] and \[ Pr[w=\ell_a] = 1- Pr[w=w_a] \]

Further, given the sample ${\bf s}_n$, $W$ inherits a {\em posterior} distribution, $Pr[W \mid {\bf S}_n = {\bf s}_n]$, also known as the {\em a posteriori} probability of $W$. The Bayesian audit estimate $\hat{w}_n({\bf s}_n)$ is the candidate that maximizes this probability (that is, the candidate for whom this value is larger), with the constraint that the probability of estimation error (the upset probability) is smaller than $\gamma$, a pre-determined quantity, $0 < \gamma < \frac{1}{2}$.

The {\em (computational) Bayesian Audit} assumes the audit draws votes without replacement and uses knowledge of $f_X$ to simulate the distribution on the unexamined votes, conditional on ${\bf s}_n$, using P\'{o}lya urns. The estimate ${\hat w}_n$ is the estimate with the largest number of wins in the simulations, provided the fraction of wins is greater than $1 - \gamma$.

The drawing of votes without replacement becomes particularly important in a tight election, where low margins tend to require large samples; an audit with replacement could result in more than $N$ draws, while an audit without replacement would definitely stop in $N$ draws. Sampling without replacement makes far more efficient use of the information in a sample, and has since been adopted in the design of traditional {\em RLA}s as well, see, for example, \cite{suite}.

The three contributions of Rivest and Shen (sampling without replacement, the Bayesian approach for election audits and computational Bayesian audits) do not have to be used together: sampling without replacement can be used for any audit, and Bayesian audits do not have to be computed using P\'{o}lya urn simulations.

We study the general Bayesian audit and do not restrict ourselves to P\'{o}lya urn simulations; this is particularly easy in the two-candidate election. We will refer to the general Bayesian audit as described above as the $(\gamma, f_X)$-Bayesian audit. Further, to avoid confusion, the term will always refer to audits without replacement. We do not explore Bayesian audits with replacement. Additionally, as mentioned earlier, we assume that $Pr[w=w_a] = Pr[w=\ell_a]$\\

\item {\em Audits with and without replacement}

So that we may understand better all the differences between Bayesian audits and other audits, we provide expressions for both the Wald Sequential Test and the {\em traditional RLA} without replacement. We note that the likelihood ratio is the ratio of the probabilities of drawing a single permutation (denoted by the specific audit sequence of $n$ draws that $s_n$ represents) when the number of votes for $w_a$ is $p_1N$ and $p_0N$ respectively.

That is,
\[ \sigma _n =  \frac{Pr[{\bf S}_n = {\bf s}_n \mid w=w_a]}{Pr[{\bf S}_n = {\bf s}_n \mid w = \ell_a] } = \frac{\frac{hg(k_{{\bf s}_n}, N, p_1N, n)}{\binom{n}{k_{{\bf s}_n}}}}{\frac{hg(k_{{\bf s}_n}, N, p_0N, n)}{\binom{n}{k_{{\bf s}_n}}}}\]
where $hg(k, N, x, n)$ is the hypergeometric distribution: the probability of obtaining $k$ items with the desired characteristic when $n$ items are drawn from a total of $N$ items of which $x$ have the desired characteristic. In our case, the items are votes in the election, and those with the desired characteristic are votes for $w_a$. Thus $hg(k_{{\bf s}_n}, N, x, n)$ is the probability of drawing $k_{{\bf s}_n}$ votes for $w_a$ in a sample of size $n$ drawn from the $N$ votes cast in the election, of which $x$ is the true number of votes for $w_a$. Dividing this value by $\binom{n}{k_{{\bf s}_n}}$ gives us the probability of drawing a particular sequence of $n$ votes of which $k_{{\bf s}_n}$ are for $w_a$. (The term $\binom{n}{k_{{\bf s}_n}}$ will be common to the numerator and denominator in the likelihood ration and will cancel out).

The $(\alpha, \beta, p_0, p_1)$ {\em Wald Sequential Test without replacement} is:
\begin{equation}
\hat{w}_n =  \left\{ \begin{array}{ll} w_a & ~~~\frac{hg(k_{{\bf s}_n}, N, p_1N, n)}{hg(k_{{\bf s}_n}, N, p_0N, n)} > \frac{1-\beta}{\alpha}\\
& \\
\ell _a& ~~~ \frac{hg(k_{{\bf s}_n}, N, p_1N, n)}{hg(k_{{\bf s}_n}, N, p_0N, n)} < \frac{\beta}{1-\alpha} \\
& \\
undetermined ~(draw~more~samples) & ~~~else \\ \\
\end{array}
\right .
\label{eqn:wald-f_X}
\end{equation}
The reader may compare the above to (\ref{eqn:wald-test}) and (\ref{eqn:wald-p0-p1}).

Similarly, the $(\alpha, \beta, p)$-{\em traditional RLA} without replacement is:
\begin{equation}
\hat{w}_n =  \left\{ \begin{array}{ll} w_a & ~~~\frac{hg(k_{{\bf s}_n}, N, pN, n)}{hg(k_{{\bf s}_n}, N, \frac{N}{2}, n)} > \frac{1-\beta}{\alpha}\\
& \\
\ell _a& ~~~ \frac{hg(k_{{\bf s}_n}, N, pN, n)}{hg(k_{{\bf s}_n}, N, \frac{N}{2}, n)} < \frac{\beta}{1-\alpha} \\
& \\
undetermined ~(draw~more~samples) & ~~~else \\ \\
\end{array}
\right .
\label{eqn:bravo-without}
\end{equation}
where $p$ is the fraction of votes declared for $w_a$. The reader may compare the above to (\ref{eqn:bravo}). Note that, for $\beta=0$, this is the $(\alpha, p)$-{\em BRAVO} audit without replacement.
\end{enumerate}

\section{Relationships Among the Audits}
\label{sec:theorems}
In this section we demonstrate relationships among the different types of audits described in the previous section. Some of the material presented is obvious, some might have been shown elsewhere. But we are not aware of this material appearing together elsewhere in similar form, and we believe Theorems 1 and 2, at the very least, are original.
\subsection{A general expression for the Bayesian audit without replacement}
In this section we derive a general expression for the Bayesian audit. \\

{\em Theorem 1:} The $(\gamma, f_X)$-Bayesian audit is of the form:
\begin{equation}
\hat{w}_n =  \left\{ \begin{array}{ll} w_a & ~~~\tau_n > \frac{1-\gamma}{\gamma} \\
& \\
\ell_a & ~~~\tau_n < \frac{\gamma}{1-\gamma} \\
& \\
undetermined ~(draw~more~samples) & ~~~else\\
\end{array}
\right .
\label{eqn:bayesian}
\end{equation}
where $\tau _n$ is the ratio:
\[ \tau _n =  \frac{Pr[w =w_a \mid {\bf S}_n = {\bf s}_n] }{Pr[w =\ell_a \mid {\bf S}_n = {\bf s}_n] } = \sigma_n \times \frac{Pr[w=w_a]}{Pr[w=\ell_a]}\]

or
\begin{equation}
\tau _n =  \frac{\sum_{x=\frac{N+1}{2}}^N hg(k_{{\bf s}_n}, n, x, N) f_X(x)}{\sum_{x=0}^{\frac{N-1}{2}} hg(k_{{\bf s}_n}, n, x, N) f_X(x)}
\label{eqn:tau}
\end{equation}

{\em Proof:} See the appendix for the proof. (Note that, for the summations in (\ref{eqn:tau}), it is $x$ that varies, while, for the hypergeometric distribution, it is $k$ that varies. A normalizing factor in both the numerator and denominator---from the application of Bayes' theorem---accounts for this, and cancels out.)\\

Readers may compare the expression of Theorem 1 with expression (\ref{eqn:bravo-without}).

\subsection{The Wald Sequential Test and the Bayesian Audit}
Some Wald Sequential Tests are instances of a Bayesian audit. In this section, we make this relationship precise. Here, $\delta_{x,a}$ denotes the (discrete) Kronecker delta function which takes on the value $1$ for $x=a$ and is zero otherwise.\\

{\em Corollary 2:} The $(\gamma, \gamma, p_0, p_1)$ Wald Sequential Test without replacement is the $(\gamma, f_X)$-Bayesian audit for
\[ f_X = \frac{1}{2} \delta_{x,p_0N} + \frac{1}{2} \delta_{x,p_1N} \]

Proof: The result follows trivially from (\ref{eqn:wald-f_X}), (\ref{eqn:bayesian}) and (\ref{eqn:tau}).\\

\subsection{The {\em traditional RLA} as a Bayesian Audit}

When the only possible values for $x$ are $p_0N$or $p_1N$, the value $\alpha$ is the risk of the $(\alpha, \beta, p_0, p_1)$ Wald Sequential Test. The risk of the test could be larger if the true value of $x$ is not one of these two. On the other hand, whatever the true value of $x$, the risk of the $(\alpha, \beta, \frac{1}{2}, p_1)$ Wald Sequential Test will not be larger than $\alpha$; that is, the $(\alpha, \beta, \frac{1}{2}, p_1)$ Wald Sequential Test is an $\alpha$-{\em RLA}. While we have not seen this result proven in the literature, it is well-known, and  related to Theorem 2, which we prove later.

The {\em traditional RLA} (\ref{eqn:bravo}) is a special case of the $(\alpha, \beta, p_0, p_1)$ Wald Sequential Test, with $p_0=\frac{1}{2}$ and $p_1$ chosen according to various considerations. From Corollary 2, when $\alpha = \beta$ the {\em traditional RLA} is also a Bayesian audit with $\gamma=\alpha=\beta$.\\

{\em Corollary 3:} The $(\gamma, \gamma, p)$-{\em traditional RLA} without replacement is the $(\gamma, f_X)$-Bayesian audit for
\[ f_X = \frac{1}{2} \delta_{x,\frac{N}{2}} + \frac{1}{2} \delta_{x,pN}\]

Proof: The result follows trivially from Corollary 2 and the fact that the $(\gamma, \gamma, p)$-{\em traditional RLA} is the $(\gamma, \gamma, \frac{1}{2}, p)$-Wald Sequential Test.\\

Note that the $(\alpha, p)$-{\em BRAVO} audit may not be represented as a special case of the above because the Bayesian audit as defined by Rivest and Shen requires $\alpha = \beta$. However, a more general definition of the Bayesian audit, where the probability of erring when the outcome is correct is zero and not equal to the probability of erring when the outcome is wrong, would correspond to the {\em BRAVO} audit for $f_X$ as above.

\subsection{A General {\em RLA}}
In this section, we see that we can define a general form of the {\em RLA} using the Bayesian model. To do so, we first examine in more detail the risk of an audit.

\subsubsection{The Risk of any Audit}
We first establish additional notation in order to represent the risk. Given any audit, consider the audit sample sequences ${\bf s}_n$ for which the audit stops with $\hat{w}_n({\bf s}_n) = w_a$. Denote this set of sample sequences by $\Lambda$. Note that $n$ is not fixed because the number of samples drawn is not fixed, thus $\Lambda$ contains sequences of different lengths. Similarly, denote by $\overline{\Lambda}$ the set of sample sequences for which the audit ends with $\hat{w}_n({\bf s}_n) = \ell_a$. \\

{\em Lemma 1:} The $(\gamma, f_X)$-Bayesian audit has
 \[ Pr[\hat{w}_n({\bf s}_n)=w_a \mid w \neq w_a] = P_M = \sum_{{\bf s}_n \in \Lambda} \frac{\sum_{x=0}^{\frac{N-1}{2}} hg(k_{{\bf s}_n}, n, x, N) f_X(x)}{\binom{n}{k_{{\bf s}_n}} Pr[w=\ell_a]} < \gamma \]
\[Pr[\hat{w}_n({\bf s}_n)=\ell_a \mid w = w_a] = P_U = \sum_{{\bf s}_n \in \overline{\Lambda}} \frac{\sum_{x=\frac{N+1}{2}}^N hg(k_{{\bf s}_n}, n, x, N) f_X(x)}{\binom{n}{k_{{\bf s}_n}} Pr[w=w_a]} < \gamma \]

{\em Proof:} See the Appendix for the proof.

Each audit we have covered (whether Bayesian, traditional risk-limiting or Wald Sequential) assumes a prior (the values of $p_0$ and $p_1$ for the Wald Sequential Test, $p$ for the {\em traditional RLA} and $f_X$ for the Bayesian audit). The choice of prior results in the sets $\Lambda$ and $\overline{\Lambda}$ of sample sequences ${\bf s}_n$ for which the audit stops---with $\hat{w}_n({\bf s}_n) = w_a$ or $\hat{w}_n({\bf s}_n) = \ell_a$, respectively. A different prior---for the same audit---would result in different sets $\Lambda$ and $\overline{\Lambda}$, which are independent of the particular election itself, or the true tally. These sets define the audit. The true risk of an audit is a function of the set $\Lambda$ and the true tally when $\ell_a$ wins (the unknown value of $x$). In order to avoid confusion, we will not denote the true risk by $P_M$, which has so far referred to an {\em ex ante} definition of the risk (one that does not take the true tally into consideration, but is derived using the assumed prior).

The true risk of the audit is the probability of drawing any of the sample sequences in $\Lambda$, when the number of votes for $w_a$ is $x$ for some unknown $x < \frac{N}{2}$. Denoting true risk, or {\em ex poste} risk, by $P_T$, we have:
\begin{equation}
P_T(\Lambda, x)  = \sum_{{\bf s}_n \in \Lambda} Pr[{\bf S}_n = {\bf s}_n \mid x; w = \ell_a] = \sum_{{\bf s}_n \in \Lambda} \frac{hg(k_{{\bf s}_n}, N, x, n)}{{\binom{n}{k_{{\bf s}_n}}}}
\label{eqn:risk-S}
\end{equation}
See Appendix for details.
\subsubsection{The Risk-Limiting Bayesian Audit}
We now describe a way to choose a set of points $\Lambda$ such that $P_T(\Lambda,x) < \alpha~~\forall~x$. That is, we describe an approach to obtaining an $\alpha$-{\em RLA}. For this purpose we first define the particular type of probability distribution on $x$.

Given a prior $f_X$ of the vote count for election $E$, define the {\em risk-maximizing distribution corresponding to $f_X$} (denoted $f_X^*$) as follows.
\begin{equation}
f_X^* =  \left\{ \begin{array}{ll} f_X(x) & ~~~ x \geq \frac{N+1}{2}\\
& \\
\frac{1}{2}& ~~~ x=\frac{N-1}{2} \\
& \\
0 & ~~~else \\
\end{array}
\right .
\label{eqn:f_X*}
\end{equation}
Note that $f_X^*$ is a valid distribution for the vote count of an election.\\

{\em Theorem 2:} The $(\alpha, f_X^*)$-Bayesian Audit is an $\alpha$-{\em RLA} with $P_U < \alpha$ for election $E$ with prior $f_X$.

{\em Proof:} See Appendix. \\

{\em Corollary 4:} The $(\alpha, \alpha, \frac{1}{2}, p)$-Wald Sequential Test, which is also the $(\alpha, \alpha, p)$-{\em traditional RLA}, is an $\alpha$-{\em RLA}.

{\em Proof:} Follows from Corollary 2 and Theorem 2.\\

A more general version of {\em Corollary 4}, for the $(\alpha, \beta, p)$-{\em traditional RLA} is generally known to be true, and can be proven as above but we are not aware of a proof in the literature on election audits.

\section{Computing {\em RLA}s and Bayesian Audits}
\label{sec:expts}
We defined a general Bayesian {\em RLA} in the previous section. The prior $f_X^*$ is not a natural fit to computing using P\'{o}lya urns, however. In this section we describe how the Bayesian {\em RLA} may be pre-computed. Pre-computation improves the computational efficiency of a Bayesian audit, no longer constraining us to the use of P\'{o}lya urn simulations. As a consequence, we are also not restricted to beta distributions for the prior. We begin with the pre-computation of the {\em traditional RLA}, which follows from a classical result by Wald \cite{wald}.

\subsection{Audits and pre-computed look-up tables}
\label{sec:precomputed}
We observe that the Wald, {\em traditional RLA}s and Bayesian audits may be defined in the form:
\begin{equation}
\hat{w}_n({\bf s}_n) =  \left\{ \begin{array}{ll} w_a & ~~~ k_{{\bf s}_n} \geq k^+\\
& \\
\ell _a& ~~~  k_{{\bf s}_n} \leq k^-\\
& \\
undetermined & ~~~else \\
 (draw~more~samples) & \\ \\
\end{array}
\right .
\end{equation}
where $k^+$ and $k^-$ are determined by the specific audit.

For example, for the {\em traditional RLA} (\ref{eqn:bravo}), $\frac{p^{k_{{\bf s}_n}} (1-p)^{n-k_{{\bf s}_n}}}{(\frac{1}{2})^n}$ is monotone increasing with $k_{{\bf s}_n}$ because $p > 1-p$, and hence
\[\frac{p^{k^*} (1-p)^{n-k{^*}}}{(\frac{1}{2})^n} > \frac{1-\beta}{\alpha} \Rightarrow  \frac{p^{k_{{\bf s}_n}} (1-p)^{n-k_{{\bf s}_n}}}{(\frac{1}{2})^n} > \frac{1-\beta}{\alpha} ~~~ \forall k_{{\bf s}_n} \geq k^* \]
Similarly,
\[\frac{p^{k^*} (1-p)^{n-k{^*}}}{(\frac{1}{2})^n} < \frac{\beta}{1-\alpha} \Rightarrow  \frac{p^{k_{{\bf s}_n}} (1-p)^{n-k_{{\bf s}_n}}}{(\frac{1}{2})^n} < \frac{\beta}{1-\alpha} ~~~ \forall k_{{\bf s}_n} \leq k^* \]

Let $k^+$ be the smallest integer $k$ such that
\[\frac{p^{k} (1-p)^{n-k}}{(\frac{1}{2})^n} > \frac{1-\beta}{\alpha} \]
and $k^-$ the largest integer $k$ such that:
\[  \frac{p^k (1-p)^{n-k}}{(\frac{1}{2})^n} < \frac{\beta}{1-\alpha} \]
and hence:
\begin{equation}
\label{eqn:kplus}
k_+ =ceiling( \frac{log(\frac{1-\beta}{\alpha})}{log (\frac{p}{1-p})} + n\frac{log(\frac{\frac{1}{2}}{1-p})}{log (\frac{p}{1-p}) })
\end{equation}
and
\begin{equation}
\label{eqn:kminus}
k_- =floor( \frac{log(\frac{\beta}{1-\alpha})}{log (\frac{p}{1-p})} + n\frac{log(\frac{\frac{1}{2}}{1-p})}{log (\frac{p}{1-p}) })
\end{equation}

For the {\em traditional RLA} without replacement, simple algebra demonstrates that $\frac{hg(k, N, p_1N, n)}{hg(k, N, \frac{N}{2}, n)}$ is monotone increasing with $k$ and $k^+$ is the smallest integer $k$ such that
\[ \frac{hg(k, N, p_1N, n)}{hg(k, N, \frac{N}{2}, n)} > \frac{1-\beta}{\alpha} \]
and $k^-$ the largest integer $k$ such that:
\[ \frac{hg(k, N, p_1N, n)}{hg(k, N, \frac{N}{2}, n)} < \frac{\beta}{1-\alpha} \]

For the Bayesian audit (\ref{eqn:bayesian}) too, one may show, as one may expect, that $\tau _n$ is monotone increasing with $k$.
$k^+$ is the smallest integer $k$ such that
\[ \frac{\sum_{x=\frac{N+1}{2}}^N hg(k, n, x, N) f_X(x)}{\sum_{x=0}^{\frac{N-1}{2}} hg(k, n, x, N) f_X(x)}  > \frac{1-\gamma}{\gamma}\]
and $k^-$ the largest integer $k$ such that:
\[ \frac{\sum_{x=\frac{N+1}{2}}^N hg(k, n, x, N) f_X(x)}{\sum_{x=0}^{\frac{N-1}{2}} hg(k, n, x, N) f_X(x)}  < \frac{\gamma}{1-\gamma }\]

\subsection{Experimental Verification}
We performed experiments to: (a) generate lookup tables using the expressions we derived for the general Bayesian audit, as described in Theorem 1 and section \ref{sec:precomputed}; (b) verify the lookup tables; (c) generate lookup tables for the Bayesian {\em RLA} we propose and verify that the maximum risk is as expected and (d) compare the number of samples required to stop the various audits.
\subsubsection{Generation of Lookup Tables for the Bayesian Audit}
We used the derived expressions for the Bayesian audit to form a look-up table for values of $k^+$ given values of $n$ (see Table \ref{table:kplus}), assuming the beta distribution prior with pseudo-counts of $0.5$ for each candidate ($f_X(x)$ is proportional to $x^{-\frac{1}{2}}(1-x)^{-\frac{1}{2}}$), an election with $N=100,000$ votes, and an escalating audit with sample sizes escalating by a factor of 2---beginning at 200 and ending at 51,200. We generated values of $k^+$ for each of the nine values of $n$ and $\gamma = 0.1, 0.05, 0.02, 0.01, 0.005, 0.002, 0.001$, see Table \ref{table:kplus}.

We used the computed values of $k^+$ to simulate 1,000 audits of a tied election with $N$ votes, and computed the fractional number of times the audit stopped (declaring the outcome correct); this is an estimate of the maximum risk of the audit over all possible values of the tally. We observe that the maximum risk is many times the upset probability. In unpublished simulations performed independently by us, Ottoboni and Rivest, the maximum risk is even larger for finer audit samples.
\begin{table}[h!]
\centering
\begin{tabular}{||l||c|c|c|c|c|c|c|c|c|c||}
\hline
{\bf Upset} & {\bf Max.} & \multicolumn{9}{c||}{{\bf Audit sample size,} $n$}\\ \cline{3-11}
{\bf Prob.} $\gamma$& {\bf Risk} & 200 & 400 & 800 & 1,600 & 3,200 & 6,400 & 12,800 & 25,600 & 51,200\\ \hline \hline
0.1    & 0.416     &  110        &  213        &  419        &  826        & 1,636        & 3,250        & 6,468       & 12,889       & 25,702\\ \hline
0.05 & 0.236       &   112        &  217       &   424       &   833       &  1,646       &  3,264       &  6,487      &  12,914      &  25,731 \\ \hline
0.02  & 0.138      &   115        &  221       &   429        &  841       &  1,658       &  3,280       &  6,509      &  12,942      &  25,763 \\ \hline
0.01   & 0.061      &   117        &  224       &   433        &  847       &  1,665       &  3,291       &  6,523      &  12,961      &  25,785\\ \hline
0.005  & 0.028       &   119        &  226       &   437        &  852       &  1,672       &  3,300       &  6,537      &  12,978      &  25,804\\ \hline
0.002   & 0.011      &   121        &  229       &   441        &  858       &  1,681       &  3,312       &  6,553      &  12,999      &  25,828\\ \hline
0.001   & 0.007      &   122        &  231       &   444        &  862       &  1,686       &  3,320       &  6,564      &  13,014      &  25,845\\ \hline
\end{tabular}
\caption{Values of $k^+$ computed for the Bayesian audit with $N=100,000$ and $f_X$ proportional to $x^{-\frac{1}{2}}(1-x)^{-\frac{1}{2}}$}
\label{table:kplus}
\end{table}
\subsubsection{Verification of the Bayesian Audit Expression}
To verify our results, we compared the maximum risk estimates obtain using the lookup tables to those computed using Rivest's public software library \cite{Bayesian-code} with 10,000 inner trials for the Bayesian simulations (10,000 simulations given an audit sample, to estimate the Bayesian posterior and the upset probability) and 10,000 outer trials (10,000 instances of the audit) on a tied election with $N=100,000$. We had access to the following data\footnote{Thanks to Ronald L. Rivest for providing the results of these simulations that he had carried out for a different purpose.}:
\begin{enumerate}
\item The final values of $k$---the number of votes for $w_a$ in the sample---and $n$---the size of the sample---for each of 10,000 instances of the Bayesian audit, for $\gamma = 0.1$ and $\gamma = 0.005$.
\item The fractional number of times the audit stopped (maximum risk) over 10,000 instances, for $\gamma = 0.1, 0.05, 0.02, $ $0.01, 0.005, 0.002, 0.001$.
\end{enumerate}

We know that the final values of $k$ ((1) above) are values for which the Bayesian audit computed with P\'{o}lya urn simulations stopped. We compare these values of $k$ with the corresponding values of $k^+$ computed by us to determine if the audit using the look-up table would also have stopped. We found that the error rate between the two audits was 0.0855 for $\gamma = 0.1$ and 0.0118 for $\gamma = 0.005$, which indicates considerable general agreement.

In both cases most of the errors occur at the sample size of $51,200$, when the Bayesian audit goes to a full hand count, while our expression predicts that it should stop. This is because the Bayesian audit, based on probabilistic simulations, may take an audit with $k_1$ votes for the winner to a full hand count, and also stop the audit for a sample with $k_2$ votes for the winner, when $k_2 < k_1$. We expect that this variance is a function of the number of inner trials. On the other hand, the behaviour of the look-up table audit is a deterministic and monotonic function of the value of $k$, and if a sample with $k_1$ votes goes to a full hand count, so would any sample with $k_2$ votes, for all $k_2 < k_1$.

We compared (2) above to our estimates of the risk reported in Table \ref{table:kplus}. Our results were similar (see Table \ref{table:risk}), and differences are likely attributable to the difference in the number of simulations (10,000 and 1,000) and in the finiteness of the number of simulations; that is, the values in both cases are simply estimates.
\begin{table}[h!]
\centering
\begin{tabular}{||l||c|c||}
\hline
{\bf Upset Probability $\gamma$} & {\bf Risk Limit of Bayesian Audit} & {\bf Risk Limit Estimated using Lookup Table} \\ \hline
0.1   & 0.4168 & 0.416 \\ \hline
0.05  & 0.2531 & 0.236 \\ \hline
0.02  & 0.1244 & 0.138 \\ \hline
0.01  & 0.0634 & 0.061 \\ \hline
0.005 & 0.0372 & 0.028 \\ \hline
0.002 & 0.0154 & 0.011 \\ \hline
0.001 & 0.0083 & 0.007 \\ \hline
\end{tabular}
\caption{Risk limit estimates for the seven-tier Bayesian audit for $N=100,000$ and $f_X$ proportional to $x^{-\frac{1}{2}}(1-x)^{-\frac{1}{2}}$}
\label{table:risk}
\end{table}

\subsubsection{Example Bayesian Risk Limiting Audits}
\begin{figure}[h!]
\includegraphics[scale=0.67, angle=-90, origin=c]{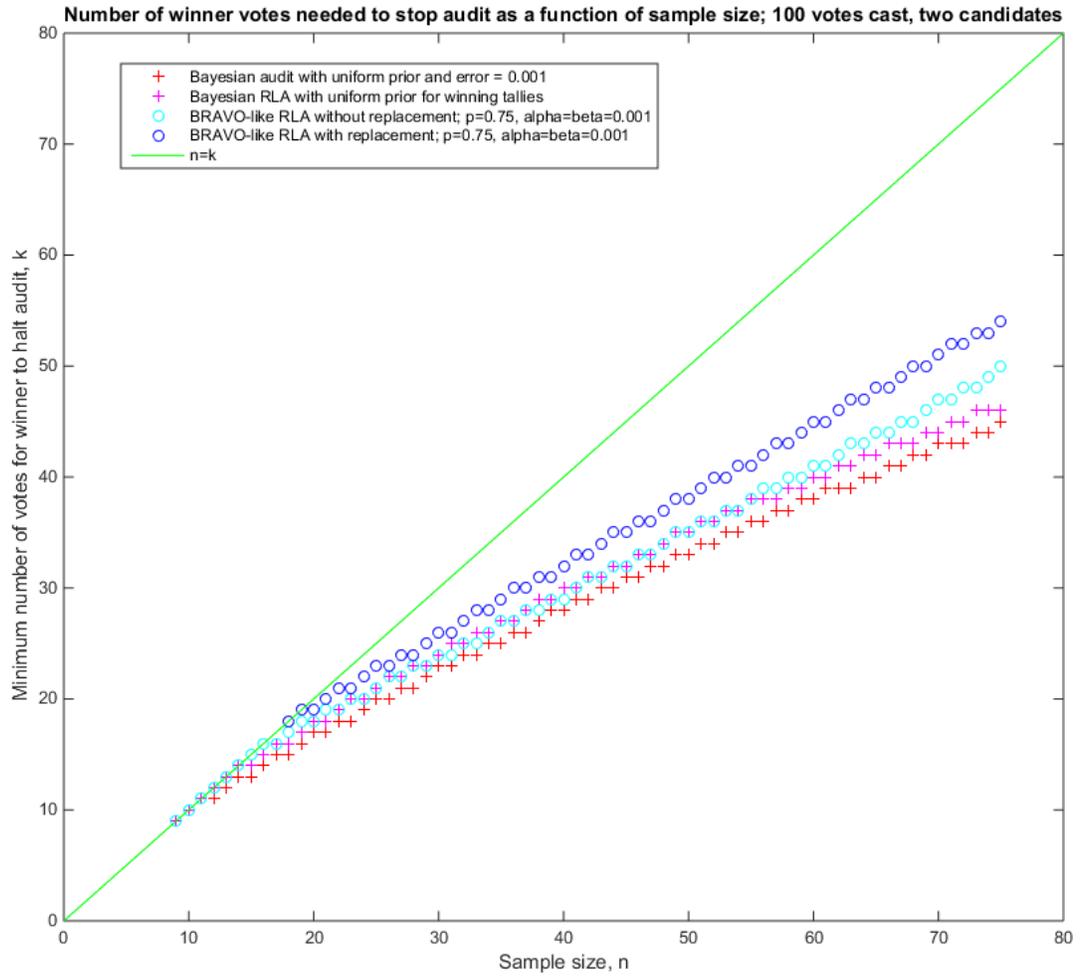}
\caption{Minimum number of winner votes as a function of sample size}
\label{fig:kplus}
\end{figure}

We computed two Bayesian RLAs.
\begin{enumerate}
\item $N=100,000$, risk measures of 0.1 and 0.005

We computed Bayesian RLAs for $N=100,000$, risk measures of 0.1, 0.05 and 0.005, and an escalating audit with sample sizes escalating by a factor of 2---beginning at 200 and ending at 51,200, and a prior that is uniform on tallies favoring the winner and concentrated on a margin of one for tallies favoring the loser. In Table \ref{table:kplus2}, we compare the values $k^+$ for the two types of Bayesian audits, standard and {\em RLA}.
\begin{table}[h!]
\centering
\begin{tabular}{||c||c||c|c|c|c|c|c|c|c|c||}
\hline
{\bf Bayesian}& Bayesian & \multicolumn{9}{c||}{{\bf Audit sample size,} $n$}\\ \cline{3-11}
{\bf error} $\gamma$& Audit Type & 200 & 400 & 800 & 1,600 & 3,200 & 6,400 & 12,800 & 25,600 & 51,200 \\ \hline \hline
0.1  & Standard     &  110        &  213        &  419        &  826        & 1,636        & 3,250        & 6,468       & 12,889       & 25,702 \\ \cline{2-11}
& RLA & 120     &    230      &   443      &   863   &     1,691    &    3,331     &   6,585    &   13,049    &   25,897 \\ \hline
0.05  & Standard     &  112        &  217       &   424       &   833       &  1,646       &  3,264       &  6,487      &  12,914      &  25,731 \\ \cline{2-11}
& RLA &  122     &    232      &   447      &   868   &     1,698    &    3,339     &   6,596    &   13,063    &   25,913 \\ \hline
0.005 & Standard & 119        &  226       &   437        &  852       &  1,672       &  3,300       &  6,537      &  12,978      &  25,804 \\ \cline{2-11}
& RLA & 127    &     239    &     456    &     880     &   1,715    &    3,363     &   6,627    &   13,103    &   25,957 \\ \hline
\end{tabular}
\caption{Values of $k^+$ computed for the Bayesian RLA with $N=100,000$ and constant $f_X$, and compared to the Bayesian audit with the same parameters}
\label{table:kplus2}
\end{table}
Because the range in the values of $k^+$ is very large, the differences between the values of $k^+$ for the two types of audits are not visible in a figure. Instead, Figure \ref{fig:differences} contains plots of the difference between $k^+$ for the Bayesian RLA and the standard Bayesian audit.
\begin{figure}[h!]
\includegraphics[scale=0.33]{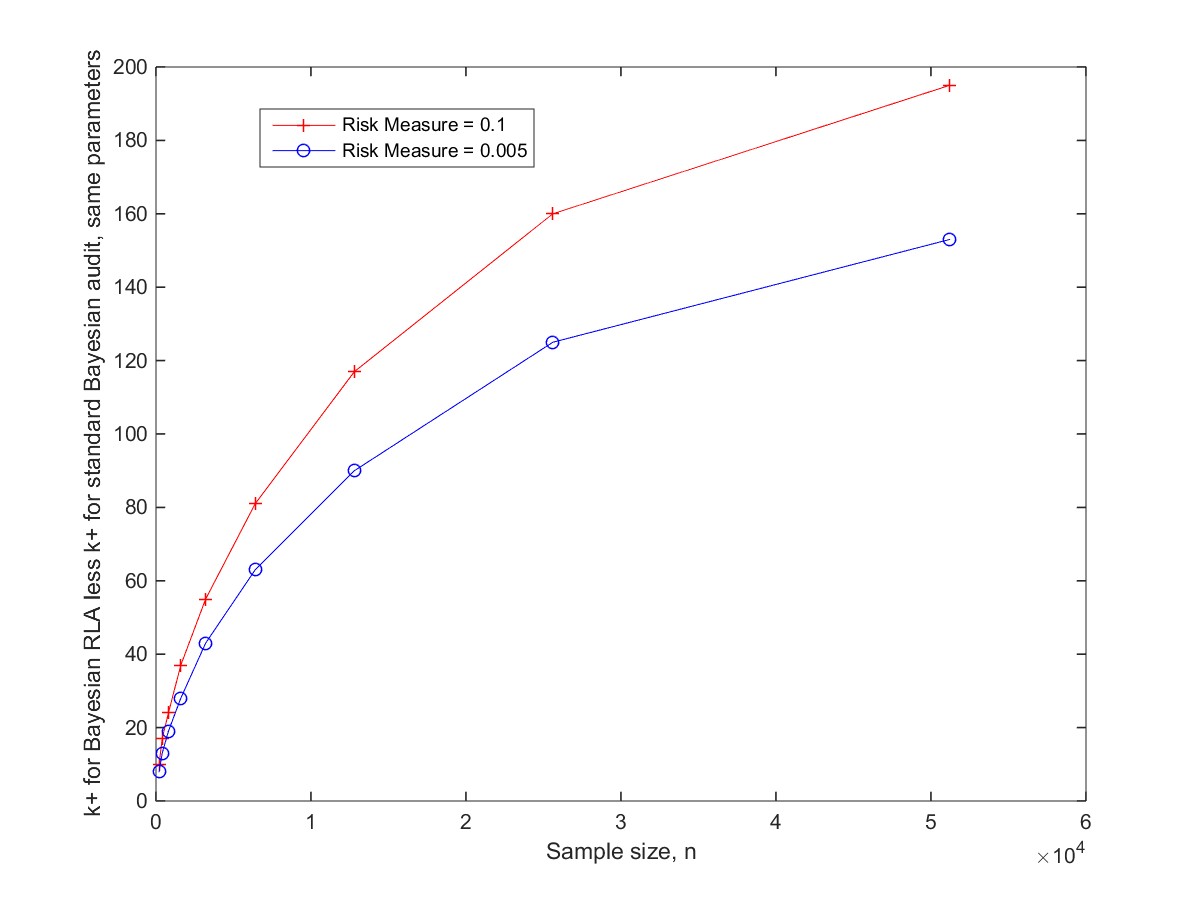}
\caption{The minimum number of winner votes for the Bayesian RLA ($k^+$) less those for the standard Bayesian audit as a function of sample size.}
\label{fig:differences}
\end{figure}
\item $N=100$, risk measure 0.001.

We computed values of $k^+$ for $N=100$, two candidates and risk measure 0.001 and audit sample sizes from 9-78 and compared the following audits:
\begin{enumerate}
\item Traditional {\em RLA} with replacement, $p=0.75$. That is, if the declared winner has won the election, we assume it is with a fractional vote count of $0.75$.
\item Traditional {\em RLA} without replacement, $p=0.75$.
\item Bayesian {\em RLA} corresponding to the uniform distribution. That is, the prior is uniform over all winning tallies, and the only possibility for $w \neq w_a$ is a fractional vote of $0.5$ (a tie), with probability 0.5. The fractional vote of $0.75$ in the traditional {\em RLA}s was chosen because the center of mass of the Bayesian prior when $w=w_a$ is a fractional vote of 0.75.
\item The Bayesian audit corresponding to the uniform distribution.
\end{enumerate}
Figure \ref{fig:kplus} plots the values of $k^+$ for samples sizes from 9 through 75. We observe that the audits as listed above are in increasing order of leniency. In particular, we note that the Bayesian {\em RLA} requires fewer votes for the winner than does the traditional {\em RLA} without replacement, which is interesting. We also notice that the traditional {\em RLA} with replacement requires the largest number of votes for the winner, and the Bayesian audit the smallest. This is as expected. Note that traditional {\em RLA}s are denoted {\em BRAVO}-like {\em RLA}s in the figure.
\end{enumerate}

\section{Conclusions and Future Work}
\label{sec:conc}
We have defined a risk-limiting Bayesian polling audit for two-candidate elections and demonstrated that any Bayesian polling audit for two-candidate elections is a simple comparison test between the number of votes for the announced winner in a sample and a pre-computed value for that sample size. Open questions include the application of this model to comparison audits and audits for more complex elections. Also open are the problems of an efficient algorithm to obtain the pre-computed values and the use of this approach in optimizing various election-related criteria.
\section{Acknowledgements}
\label{sec:ack}
The author gratefully acknowledges conversations with Mark Lindeman, Ronald L. Rivest and Philip B. Stark, and some experimentation by Shiva Omrani. This work was supported, in part, by NSF Award 1421373. 
\bibliography{audits}
\section{Appendix}
{\em Theorem 1:} The $(\gamma, f_X)$-Bayesian audit is of the form:
\begin{equation*}
\hat{w}_n =  \left\{ \begin{array}{ll} w_a & ~~~\tau_n > \frac{1-\gamma}{\gamma} \\
& \\
\ell_a & ~~~\tau_n < \frac{\gamma}{1-\gamma} \\
& \\
undetermined ~(draw~more~samples) & ~~~else\\
\end{array}
\right .
\end{equation*}
where $\tau _n$ is the ratio:
\[ \tau _n =  \frac{Pr[w =w_a \mid {\bf S}_n = {\bf s}_n] }{Pr[w =\ell_a \mid {\bf S}_n = {\bf s}_n] } = \sigma_n \times \frac{Pr[w=w_a]}{Pr[w=\ell_a]}\]

or
\begin{equation*}
\tau _n =  \frac{\sum_{x=\frac{N+1}{2}}^N hg(k_{{\bf s}_n}, n, x, N) f_X(x)}{\sum_{x=0}^{\frac{N-1}{2}} hg(k_{{\bf s}_n}, n, x, N) f_X(x)}
\end{equation*}

Further, $P_M, P_U < \gamma$, and:
\[ P_M = \sum_{{\bf s}_n \in \Lambda} \frac{\sum_{x=0}^{\frac{N-1}{2}} hg(k_{{\bf s}_n}, n, x, N) f_X(x)}{\binom{n}{k_{{\bf s}_n}} Pr[w=\ell_a]} \]
\[ P_U = \sum_{{\bf s}_n \in \overline{\Lambda}} \frac{\sum_{x=\frac{N+1}{2}}^N hg(k_{{\bf s}_n}, n, x, N) f_X(x)}{\binom{n}{k_{{\bf s}_n}} Pr[w=w_a]} \]
\\

Proof: The audit stops when the estimation error is smaller than $\gamma$; because this is a binary election, it stops with $\hat{w}_n({\bf s}_n) =w_a$ when:
\[ Pr[w =\ell_a \mid {\bf S}_n = {\bf s}_n] < \gamma \]
and, hence, when:

\[ Pr[w = w_a \mid {\bf S}_n = {\bf s}_n] > 1-\gamma \]

Thus $\hat{w}_n({\bf s}_n) = w_a$ when:
\[ \frac{Pr[w =w_a \mid {\bf S}_n = {\bf s}_n] }{Pr[w =\ell_a \mid {\bf S}_n = {\bf s}_n] }  > \frac{1-\gamma}{\gamma} \]

Similarly, $\hat{w}_n({\bf s}_n) =\ell_a$ when:
\[ \frac{Pr[w =w_a \mid {\bf S}_n = {\bf s}_n] }{Pr[w =\ell_a \mid {\bf S}_n = {\bf s}_n] }  < \frac{\gamma}{1-\gamma} \]

Hence the Bayesian Audit is of the form:
\begin{equation*}
\hat{w}_n =  \left\{ \begin{array}{ll} w_a & ~~~\tau_n > \frac{1-\gamma}{\gamma} \\
& \\
\ell_a & ~~~\tau_n < \frac{\gamma}{1-\gamma} \\
& \\
undetermined ~(draw~more~samples) & ~~~else\\
\end{array}
\right .
\end{equation*}
where
 $\tau _n$ is the ratio:
\[ \tau _n = \frac{Pr[w =w_a \mid {\bf S}_n = {\bf s}_n] }{Pr[w =\ell_a \mid {\bf S}_n = {\bf s}_n] }  = \frac{Pr[w =w_a \mid {\bf S}_n = {\bf s}_n] \times Pr[{\bf S}_n = {\bf s}_n] }{Pr[w =\ell_a \mid {\bf S}_n = {\bf s}_n] \times Pr[{\bf S}_n = {\bf s}_n]}
=  \frac{Pr[{\bf S}_n = {\bf s}_n \mid w=w_a] \times Pr[w=w_a]}{Pr[{\bf S}_n = {\bf s}_n \mid w=\ell_a] \times Pr[w=\ell_a]} \]
\[ = \sigma_n \times \frac{Pr[w=w_a]}{Pr[w=\ell_a]}\]

Continuing, we observe:
\[\tau _n =  \frac{Pr[w =w_a \mid {\bf S}_n = {\bf s}_n] }{Pr[w =\ell_a \mid {\bf S}_n = {\bf s}_n] } =  \frac{Pr[w =w_a \mid {\bf S}_n = {\bf s}_n] \times Pr[{\bf S}_n = {\bf s}_n] }{Pr[w =\ell_a \mid {\bf S}_n = {\bf s}_n] \times Pr[{\bf S}_n = {\bf s}_n]} = \frac{Pr[w =w_a; {\bf S}_n = {\bf s}_n]}{Pr[w =\ell_a; {\bf S}_n = {\bf s}_n]} \]
Which is, further:
\[ \frac{Pr[w=w_a; {\bf S}_n = {\bf s}_n] }{Pr[w =\ell_a; {\bf S}_n = {\bf s}_n] } = \frac{\sum_{x=\frac{N+1}{2}}^N Pr[X=x; {\bf S}_n = {\bf s}_n] }{\sum_{x=0}^{\frac{N-1}{2}} Pr[X=x; {\bf S}_n = {\bf s}_n] } =  \frac{\sum_{x=\frac{N+1}{2}}^N Pr[{\bf S}_n = {\bf s}_n \mid X=x] Pr[X=x]}{\sum_{x=0}^{\frac{N-1}{2}} Pr[{\bf S}_n = {\bf s}_n \mid X=x] Pr[X=x]} \]
\[\]
\[ = \frac{\sum_{x=\frac{N+1}{2}}^N \frac{hg(k_{{\bf s}_n}, N, x, n) f_X(x)}{\binom{n}{k_{{\bf s}_n}}}}{\sum_{x=0}^{\frac{N-1}{2}} \frac{hg(k_{{\bf s}_n}, N, x, n) f_X(x)}{\binom{n}{k_{{\bf s}_n}}}} = \frac{\sum_{x=\frac{N+1}{2}}^N hg(k_{{\bf s}_n}, N, x, n) f_X(x)}{\sum_{x=0}^{\frac{N-1}{2}} hg(k_{{\bf s}_n}, N, x, n) f_X(x)} \]

{\em Lemma 1:} The $(\gamma, f_X)$-Bayesian audit has
 \[ P_M = \sum_{{\bf s}_n \in \Lambda} \frac{\sum_{x=0}^{\frac{N-1}{2}} hg(k_{{\bf s}_n}, n, x, N) f_X(x)}{\binom{n}{k_{{\bf s}_n}} Pr[w=\ell_a]} < \gamma \]
\[ P_U = \sum_{{\bf s}_n \in \overline{\Lambda}} \frac{\sum_{x=\frac{N+1}{2}}^N hg(k_{{\bf s}_n}, n, x, N) f_X(x)}{\binom{n}{k_{{\bf s}_n}} Pr[w=w_a]} < \gamma \]

{\em Proof:}
We now show that $P_M, P_U < \gamma$, using an approach very similar to the proof of Proposition 1 by Wald. For ${\bf s}_n \in \Lambda$,
\[ Pr[{\bf S}_n = {\bf s}_n \mid w = \ell_a] = \frac{Pr[w = \ell_a \mid {\bf S}_n = {\bf s}_n] \times Pr[{\bf S}_n = {\bf s}_n]}{Pr[w=\ell_a]} \]
\[= \frac{Pr[w = \ell_a \mid {\bf S}_n = {\bf s}_n] \times (Pr[{\bf S}_n = {\bf s}_n \mid w = \ell_a] Pr[w=\ell_a] + Pr[{\bf S}_n = {\bf s}_n \mid w = w_a] Pr[w=w_a] )}{Pr[w=\ell_a]} \]
\[= Pr[w = \ell_a \mid {\bf S}_n = {\bf s}_n] \times (Pr[{\bf S}_n = {\bf s}_n \mid w = \ell_a] + Pr[{\bf S}_n = {\bf s}_n \mid w = w_a] )\]
\[< \gamma (Pr[{\bf S}_n = {\bf s}_n \mid w = \ell_a] + Pr[{\bf S}_n = {\bf s}_n \mid w = w_a] )\]
Hence,
\[ P_M = \sum_{{\bf s}_n \in \Lambda} Pr[{\bf S}_n = {\bf s}_n \mid w = \ell_a] < \gamma P_M + \gamma \sum_{{\bf s}_n \in \Lambda} Pr[{\bf S}_n = {\bf s}_n \mid w = w_a] = \gamma P_M + \gamma (1-P_U)  \]

Similarly,
\[ P_U < \gamma P_U + \gamma (1-P_M)\]
This gives us $P_M + P_U < 2\gamma$, and hence at least one of $P_M$ and $P_U$ is smaller than $\gamma$; also each is smaller than $2\gamma$.

Further,
\[ \frac{P_U}{1-P_M} < \frac{\gamma}{1-\gamma} \]
and
\[ \frac{P_M}{1-P_U} < \frac{\gamma}{1-\gamma} \]
For small $\gamma$, both $P_U$ and $P_M$ are small and $1-P_M, 1-P_U, 1-\gamma, \approx 1$. Hence $P_M < \gamma $ and $P_U < \gamma$ is approximately true.

Finally:
\[ P_M = \sum_{{\bf s}_n \in \Lambda} Pr[{\bf S}_n = {\bf s}_n \mid w = \ell_a] = \sum_{{\bf s}_n \in \Lambda}\frac{Pr[w=\ell_a; {\bf S}_n = {\bf s}_n] }{Pr[w=\ell_a]} =\sum_{{\bf s}_n \in \Lambda} \frac{\sum_{x=0}^{\frac{N-1}{2}} hg(k_{{\bf s}_n}, N, x, n) f_X(x)}{\binom{n}{k_{{\bf s}_n}} Pr[w=\ell_a]} \]
\[ P_U = \sum_{{\bf s}_n \in \overline{\Lambda}} \frac{Pr[w=w_a; {\bf S}_n = {\bf s}_n] }{Pr[w=w_a]} = \sum_{{\bf s}_n \in \overline{\Lambda}} \frac{\sum_{x=\frac{N+1}{2}}^N hg(k_{{\bf s}_n}, N, x, n) f_X(x)}{\binom{n}{k_{{\bf s}_n}} Pr[w=w_a]} \]

{\em Details for Equation (\ref{eqn:risk-S})}:
\begin{equation*}
P_T(\Lambda, x)  = \sum_{{\bf s}_n \in \Lambda} Pr[{\bf S}_n = {\bf s}_n \mid x; w = \ell_a] = \sum_{{\bf s}_n \in \Lambda} \frac{hg(k_{{\bf s}_n}, N, x, n)}{\binom{n}{k_{{\bf s}_n}}}
\end{equation*}

Proof: Looking at all sequences of length $N$,
\[ P_T(\Lambda, x)  = \sum_{{\bf s}_n \in \Lambda}~~~~ \sum_{c_{N-n} \in \{w_a, \ell_a\}^{N-n}}Pr[({\bf s}_n \mid \mid c_{N-n}) \mid x; w = \ell_a] = \sum_{{\bf s}_n \in \Lambda} Pr[{\bf S}_n = {\bf s}_n \mid x; w = \ell_a]\]
\[ = \sum_{{\bf s}_n \in \Lambda} \frac{hg(k_{{\bf s}_n}, N, x, n)}{\binom{n}{k_{{\bf s}_n}}} \]\\

{\em Theorem 2:} The $(\alpha, f_X^*)$-Bayesian Audit is an $\alpha$-{\em RLA} with $P_U < \alpha$ for election $E$ with prior $f_X$.\\

{\em Proof}:
By Theorem 1, for the $(\alpha, f_X^*)$-Bayesian Audit
\[ P_M = \sum_{{\bf s}_n \in \Lambda} \frac{hg(k_{{\bf s}_n}, N, \frac{N-1}{2}, n)\times f_X^*(\frac{N-1}{2})}{\binom{n}{k_{{\bf s}_n}}Pr[w=\ell_a]}  = \sum_{{\bf s}_n \in \Lambda} \frac{hg(k_{{\bf s}_n}, N, \frac{N-1}{2}, n)}{\binom{n}{k_{{\bf s}_n}}} < \alpha \]
and
\[ P_U = \sum_{{\bf s}_n \in \overline{\Lambda}} \frac{\sum_{x=\frac{N+1}{2}}^N hg(k_{{\bf s}_n}, n, x, N) f_X(x)}{\binom{n}{k_{{\bf s}_n}} Pr[w=w_a]} < \alpha \]

Further, using simple algebra, one can show that, for fixed $N$, $n$ and $k \in (\frac{n}{2}, n]$, $hg(k, N, x, n)$ is a monotone increasing function of $x$ for $x \in [0, \frac{N-1}{2}]$. That is,
\begin{equation*}
k > \frac{n}{2} \Rightarrow hg(k, N, x, n) \leq hg(k, N, \frac{N-1}{2}, n)~~\forall x \in [0, \frac{N-1}{2}]
\end{equation*}
Additionally, for ${\bf s}_n \in \Lambda$ we have:
\[ \frac{\sum_{x=\frac{N+1}{2}}^N hg(k_{{\bf s}_n}, N, x, n) f_X^*(x)}{hg(k_{{\bf s}_n}, N, \frac{N-1}{2}, n) f_X^*(\frac{N-1}{2})} > \frac{1-\gamma}{\gamma} > 1 \]
and, again using simple algebra, one can show that, for fixed $N$, $n$ and $k \in [0, \frac{n}{2}]$, $hg(k, N, x, n)$ is a monotone decreasing function of $x$ for $x \in [\frac{N-1}{2}, N]$. That is,
\begin{equation*}
k \leq \frac{n}{2} \Rightarrow hg(k, N, x, n) \leq hg(k, N, \frac{N-1}{2}, n)~~\forall x \in [\frac{N-1}{2}, N]
\end{equation*}
Thus, for ${\bf s}_n \in \Lambda$,
\begin{equation*}
k_{{\bf s}_n} \leq \frac{n}{2} \Rightarrow hg({\bf s}_n, N, x, n) \leq hg({\bf s}_n, N, \frac{N-1}{2}, n)~~\forall x \in [\frac{N-1}{2}, N]
\end{equation*}
and:
\[ \sum_{x=\frac{N+1}{2}}^N hg(k_{{\bf s}_n}, N, x, n) f_X^*(x) \leq hg({\bf s}_n, N, \frac{N-1}{2}, n) f_X^*(\frac{N-1}{2})\]
Because this contradicts the requirement for ${\bf s}_n \in \Lambda$, $ k_{{\bf s}_n} \geq \frac{n}{2}$.

For an election with true tally $x$ and the $(\alpha, f_X^*)$-Bayesian Audit,
\begin{equation*}
P_T(\Lambda, x) = \sum_{{\bf s}_n \in \Lambda} \frac{hg(k_{{\bf s}_n}, N, x, n)}{\binom{n}{k_{{\bf s}_n}}} \leq
\sum_{{\bf s}_n \in \Lambda} \frac{hg(k_{{\bf s}_n}, N, \frac{N-1}{2}, n)}{\binom{n}{k_{{\bf s}_n}}} = P_M < \alpha ~~\forall x \in [0, \frac{N-1}{2}]
\end{equation*}
and the $(\alpha, f_X^*)$-Bayesian Audit is an $\alpha$-{\em RLA} with $P_U < \alpha$.

\end{document}